\documentclass[preprint]{revtex4-1}

\usepackage{mathtools}

\usepackage{stmaryrd}
\usepackage{amsmath}
\usepackage{amssymb}
\usepackage{graphicx}
\usepackage{bm}
\usepackage{color}
\usepackage{parskip}
\usepackage{nomencl}

\begin{document}
\title{Helical Topological Edge States in a Quadrupole Phase}

\author{Feng Liu$^1$, Hai-Yao Deng$^{2}$, and Katsunori Wakabayashi$^{1}$}
\affiliation{$^1$Department of Nanotechnology for Sustainable Energy, School of Science and Technology,
Kwansei Gakuin University, Gakuen 2-1, Sanda 669-1337, Japan}
\affiliation{$^2$School of Physics, University of Exeter, Stocker Road EX4 4QL Exeter, United Kingdom}

\begin{abstract}
Topological electric quadrupole is a recently proposed concept that
extends the theory of electric polarization of crystals to higher
orders. Such a quadrupole phase supports topological states localized
on both edges and corners. In this work, we show that in a quadrupole
phase of honeycomb lattice, topological helical edge states and pseudo-spin-polarized corner states appear by making use of
a pseudo-spin degree of freedom related to point group
symmetry. Furthermore, we argue that a general condition for emergence
of helical edge states in a (pseudo-)spinful quadrupole phase is mirror or time-reversal symmetry. 
Our results offers a way of generating topological helical edge states without spin-orbital couplings.
\end{abstract}

\maketitle

The concept of topology in electronic materials has offered us a
unique dimension to design materials with useful
properties~\cite{Klitzing1980, Haldane1988,Hasan2010}.
Especially in topological insulators (TIs), a dissipationless spin current
flows along the edges of a strip in the absence of charge
current~\cite{Kane2005,Bernevig2006}.
These topological helical edge states have potential applications in
low-power electronics~\cite{Konig2008}.
Realization of topological helical edge states usually requires a spin-orbital
coupling. How to realize topological helical edge
states without spin-orbit couplings remains as a fundamental open question~\cite{Fu2011,Alexandradinata2014}.

The recently proposed topological electric multipoles such as dipoles and
quadrupoles offer us a nice opportunity
to attack this open question~\cite{Benalcazar2017, Benalcazar2017B}.
Topological electric multipole is a generalization of the modern theory
of charge polarization to high dimensions~\cite{Zak1989,Marzari2012,Fang2012a}, which introduces a
new class of topological materials dubbed as high order TIs~\cite{Song2017, Max2018, Zhu2018, Fukui2018, Xie2018}. When
a sample with finite topological electric dipole moment is terminated with an edge,
topologically protected fractional charge will appear on the
edge~\cite{King1993, Resta1994, Zhou2015,Ota2018}. Analogously, a
finite quadrupole upon being terminated develops both topological edge and corner
states. Experiments have observed these topological corner states in various
systems such as photonic, acoustic crystals and circuit
arrays~\cite{Peterson2018, Serra-Garcia2018, Imhof2018}. Remarkably,
emergences of finite topological dipole and quadrupole do not require
spin-orbital couplings~\cite{Liu2017,Liu2018}.

In previous studies of topological electric multipole phase, (pseudo-)spin degrees
of freedom have not been paid attention so much. Without
(pseudo-)spins, electric-multipole-induced edge states
are topologically protected but not helical. These edge states suffer
from dissipation during propagation. To overcome this shorthand and gain fundamental understanding of topological 
electric multipoles, we introduce pseudo-spin degree of freedom
related to point group symmetry in a topological quadrupole phase. For concreteness, we consider a honeycomb lattice
with Kekul\'{e}-like hopping textures.  Based on this model, we argue 
that
a general condition for emergences of
 topological helical edge states in a
(pseudo-)spinful quadrupole phase is either mirror or time-reversal symmetry.

Before going into the details of the honeycomb lattice model, let us
introduce the topological electric multipoles such as dipoles and
quadrupoles first. In crystalline systems, electric multipoles are
related to Berry connection in momentum space. For example,
dipole moment in a two-dimensional (2D) system
can be expressed  as
\begin{equation}
P_i(k) = \frac{e}{|\mathcal{P}|}\sum^{N_\text{occ}}_n\int_\mathcal{P} \mathbf{A}^{n}(k,k^\prime) \cdot \mathbf{n}_i dk^\prime,  \label{P}
\end{equation}
where the summation is taken for all the occupied energy bands, the
integration is along a straight path $\mathcal{P}$ that connects two
equivalent $\mathbf{k}$ points in momentum space.
$\mathbf{n}_i$ is a unit vector along $i$-direction, and
$\mathbf{A}^{n}=i\langle u_{n\mathbf{k}}|\partial_\mathbf{k}|u_{n\mathbf{k}}\rangle$ is Berry
connection with $|u_{n\mathbf{k}}\rangle$ the periodic part of Bloch state of $n$-th
energy band. Due to gauge freedom, dipole moment is well defined up to a lattice constant.
Since there are two independent directions in 2D system, 
the dipole moment of $n$-th energy band is written as
$\mathbf{P}^n=(P^n_i,P^n_j)$ in general. 
Such the independent components of a dipole moment allow us to define 
a quadrupole as 
\begin{equation}
Q_{ij} =\sum^{N_\text{occ}}_n P^n_iP^n_j/e \label{Q}.
\end{equation}
Equation (\ref{Q}) clearly states that 
a corner state appears when both of $P^n_i$ and $P^n_j$ are not zero.
The derivation of Eqs.(\ref{P}) and (\ref{Q}) is given in Sec.~A of Supplementary Information.

\begin{figure}[t]
\begin{center}
\leavevmode
\includegraphics[clip=true,width=0.7\columnwidth]{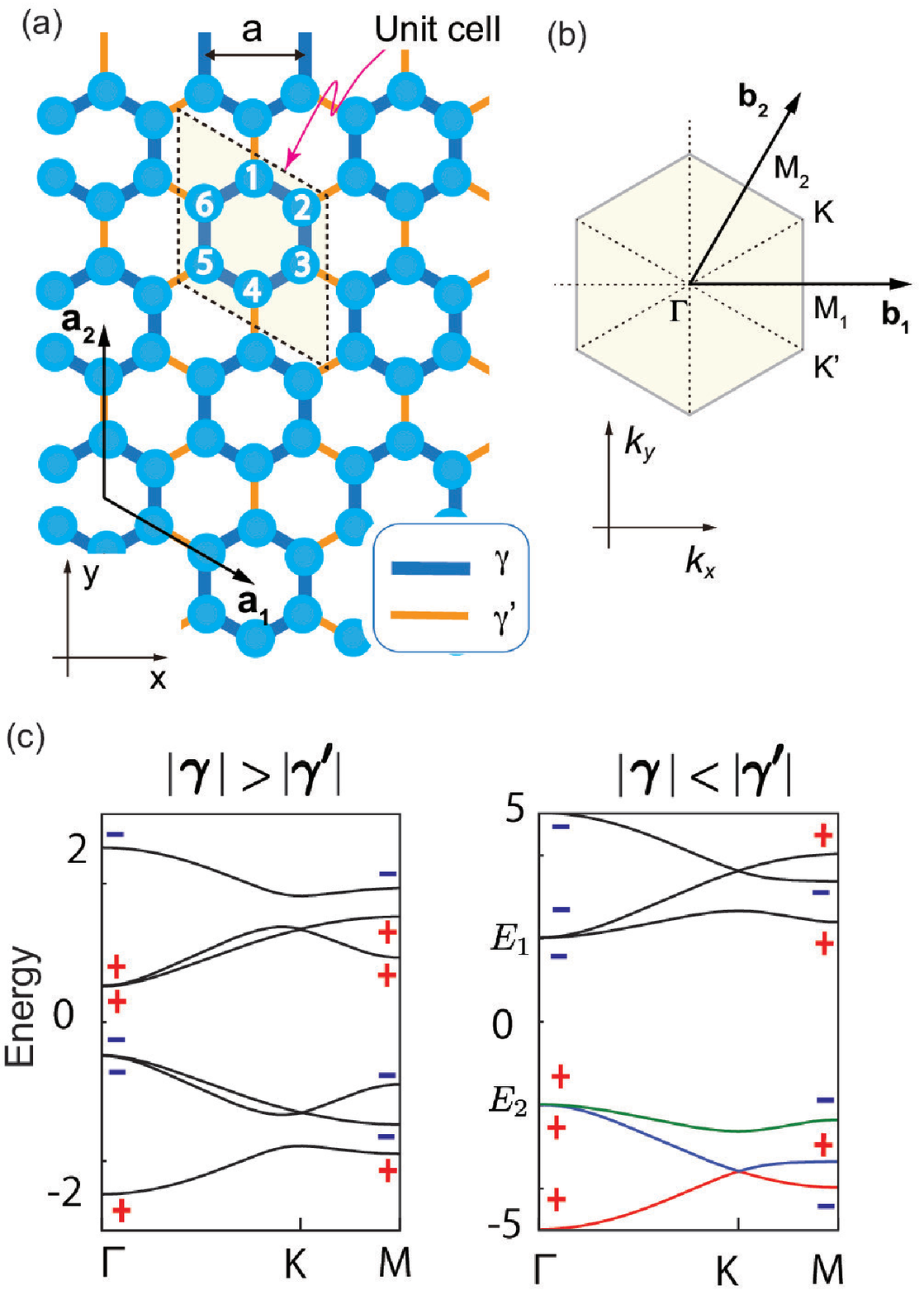}
\caption{(a) Schematic of the model which is characterized by two
  hopping parameters $\gamma$ and $\gamma'$. Within a unit cell, there
  are six atomic orbitals $|j\rangle$, $j=1,...,6$.
  (b) Reciprocal lattice vectors $\mathbf{b}_{1,2}$ and the first Brillouin zone.
  (c) Energy bands spectrum for $|\gamma|>|\gamma^\prime|$ and $|\gamma|<|\gamma^\prime|$.
  ``$\pm$'' indicates parities of wavefunctions at $\Gamma$ and
  M. M refers to either $M_1$ or $M_2$ in 1st Brillouin
  zone. First three energy bands in the case of
  $|\gamma|<|\gamma|^\prime$ are colored as red, blue and green,
  respectively, for clarifying dipole moment of each band.}
\end{center}
\end{figure}

\begin{figure}[t]
\begin{center}
\leavevmode
\includegraphics[clip=true,width=0.9\columnwidth]{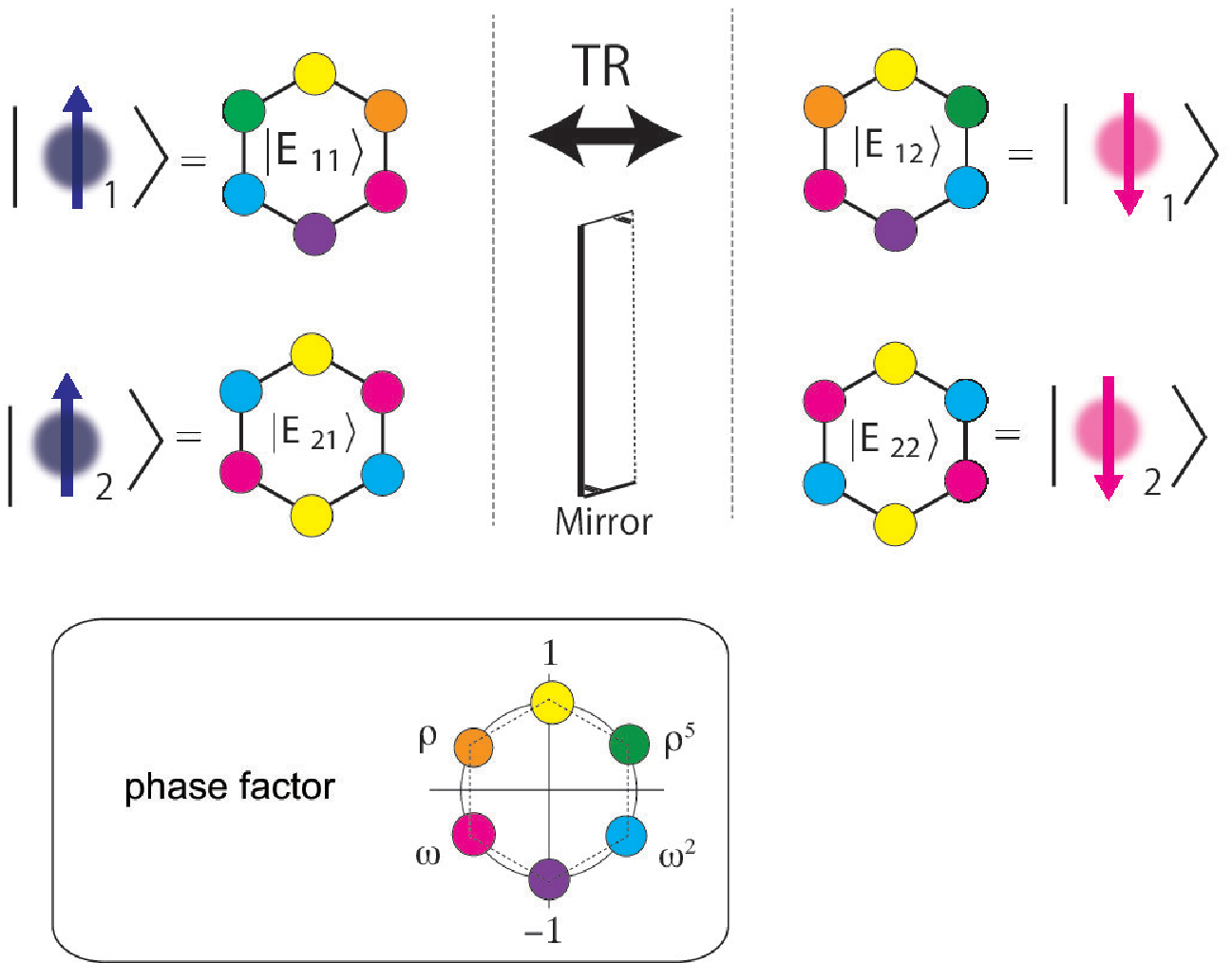}
\caption{ Schematic of pseudo-spins made up by linear combination of atomic orbitals.
  Pseudo-spin up and down transform to each by either time-reversal operation or mirror reflection.
  Inset: combination factors indicated in color.}
\end{center}
\end{figure}

The honeycomb lattice with a Kekul\'{e}-like hopping
texture is displayed in Fig.~1(a)~\cite{Kariyado2017,Liu2017B}.
There are  two types of hopping parameters
such as intra-cell hopping $\gamma$ and inter-cell hopping
$\gamma^\prime$ similar to Su-Schrieffer-Heeger (SSH) model~\cite{Delplace2011,Sorella2018}.
Resembling SSH model, topological dipole appears when
$|\gamma|<|\gamma^\prime|$~\cite{Liu2017B}.
Because of $C_{6v}$ point group symmetry that dipole moment is not zero for both $\mathbf{b}_1$ and $\mathbf{b}_2$ directions,
a finite quadrupole may exist. Here $\mathbf{b}_1$ and $\mathbf{b}_2$
are the primitive lattice vectors of the reciprocal lattice, which
form the 1st Brillouin zone as shown in Fig.~1(b). The energy band spectrums for $|\gamma|>|\gamma^\prime|$ and $|\gamma|<|\gamma^\prime|$ are displayed
in Fig.~1(c). In the case of $|\gamma|<|\gamma^\prime|$, band
inversions happen at $\Gamma$ and $M$.  A detailed energy bands evolution by changing
$\gamma/\gamma^\prime$ is given is Sec.~B of Supplementary Information. 

Since numerical evaluation of $\mathbf{A}^{n}(k,k^\prime)$ produces very spiky function in momentum space due to the gauge freedom of wavefunctions, 
it is difficult to obtain dipole moment numerically using Eq.~(1). However,
under a zero Berry curvature
($\mathcal{F}=\partial_iA_j-\partial_jA_i$)~\cite{onsites}, based on Eq.~(1) the dipole moment can be
determined by the parities at inversion-invariant
$\mathbf{k}$ points such as~\cite{Fang2012a, Liu2017B}
\begin{equation}
   P^n_i=\dfrac{\eta^n(\Gamma)}{\eta^n({M}_i)},
 \end{equation}
where $\eta^n(\mathbf{k})$ is the eigenvalue of $\pi$ rotation over
z-axis at $\mathbf{k}$ point for the $n$-th energy band. Then based the parities at $\Gamma$ and
$M_i$ shown in Fig.~1(c) for $|\gamma|<|\gamma^\prime|$, we obtain dipole moments $e/2$, $0$, $e/2$
for the 1st (red), 2nd (blue) and 3rd (green) occupied
bands, respectively.  Similarly, the quadrupoles are $e/4$, $0$ and
$e/4$, respectively, as $P^n_1=P^n_2$ guaranteed by $C_{6v}$
point group symmetry. From Eqs.~(1) and (2) the total dipole moment
vanishes and the total quadrupole is $e/2$. Thus, topological edge and
corner states appear owing to the finite quadrupole $e/2$ when
$|\gamma|<|\gamma^\prime|$. In the following we show that by introducing a pseudo-spin degree of freedom related to
$C_{6v}$ point group symmetry, topological helical edge states and
pseudo-spin-polarized corner states appear.

As shown in Fig.~1(c), owing to $C_{6v}$ point group symmetry, there are two pairs of doubly degenerate states at $\Gamma$, i.e., $E_1$ and
$E_2$ that may be regarded as pseudo-spins. For $E_1$ states, we
call them $|E_{11}\rangle$ and $|E_{12}\rangle$, which are given by
$\left|E_{11}\right>,\left|E_{12}\right>=\sum_je^{\pm\text{i}(j-1)\rho}\left|j\right>$. Similarly,
$\left|E_{21}\right>,\left|E_{22}\right>=\sum_je^{\pm\text{i}(j-1)\omega}\left|j\right>$. Here $|j\rangle$ ($j=1,...,6$) indicates six atomic orbitals in a unit
cell as indexed in Fig.~1(a), $\rho=\pi/6$ and $\omega=\pi/3$. As $E_{i1}$ and $E_{i2}$ transform into each other
under mirror reflection and also time reversal similar to real spins as depicted in Fig.~2,
we regard them as pseudo-spin degree of freedom defined as
\begin{equation}
\begin{split}
&\left|1_\uparrow\right>=\left|E_{11}\right>, \text{       } \left|2_\uparrow\right>=\left|E_{21}\right>, \\
&\left|1_\downarrow\right>=\left|E_{12}\right>, \text{       } \left|2_\downarrow\right>=\left|E_{22}\right>.
\end{split}
\end{equation}
In our model, there is no difference between pseudo-spins and real spins, 
we simply call spins from now on.

\begin{figure*}[ht]
\leavevmode
\includegraphics[clip=true,width=0.95\textwidth]{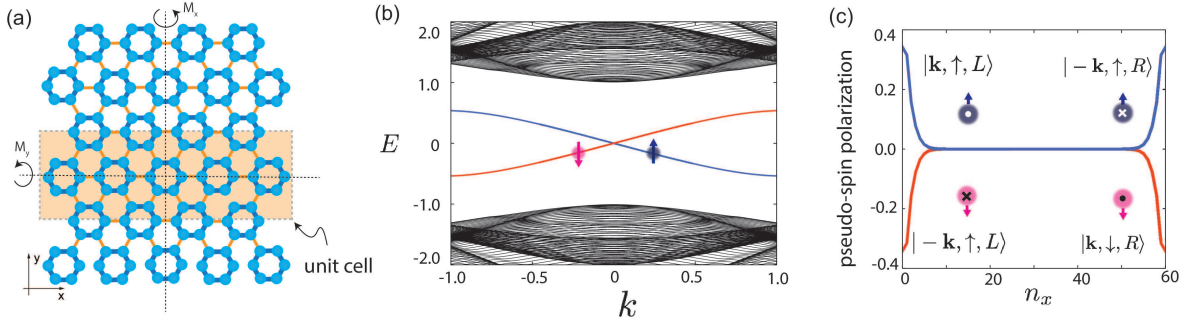}
\caption{(a) Schematic of a ribbon supporting topological helical edge states. The ribbon
  is periodic along the $y-$direction. There are two mirror planes
  denoted as
  $M_x$ and $M_y$. (b) Energy spectrum of the ribbon
  for $\gamma = -1.0$ and $\gamma' = -2.0$. The wavenumber $k$ refers
  to direction $\mathbf{b}_1$. Within the bulk energy gap, a pair of spin-polarized bands consisting of edge states
  appear. (c) Helical edge states. Due to mirror symmetry (time
  reversal symmetry), edge states of same energy must propagate oppositely -- as indicated by the cross and dot -- with opposite spin-polarization on the same edge.}
\end{figure*}

\begin{figure*}[ht]
\begin{center}
\leavevmode
\includegraphics[clip=true,width=0.95\textwidth]{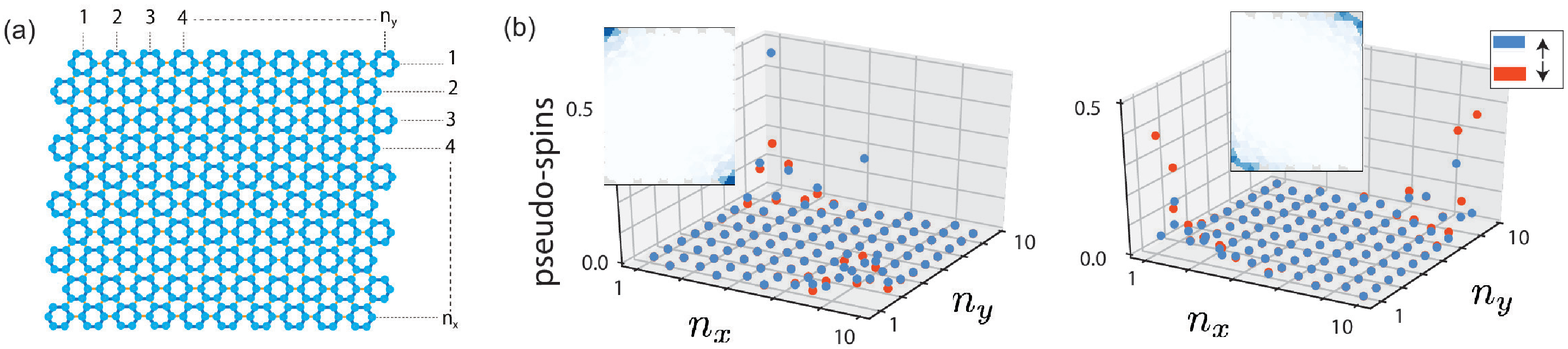}
\caption{(a) Schematic of a sample comprised of $10\times10$ unit
  cells supporting the corner states. We call the edge along the
  $x$-direction zigzag, while that along the $y$-direction
  armchair. (b) Spin-decomposition of the corner states with positive energies, blue (red)
  for the  spin-up (-down) component. Zigzag corner states are
  spin-polarized up while armchair corner states are spin-polarized down. Insets: charge density maps for the corner states. }
\end{center}
\end{figure*}

To demonstrate the existence of helical edge states, we consider a ribbon structure extended along the
$\mathbf{a}_2$ direction as displayed in Fig.~3(a). By solving its corresponding Hamiltonian, we obtain the energy spectrum of
the ribbon as displayed in Fig.~3(b). It is clear to see that there is a pair of edge-state energy bands
appearing within the band gap in Fig.~3(b).
To show that these edge states are helical, we calculate the spin-polarization defined by
$\text{SP}=|\langle \alpha_\uparrow | \psi \rangle |^2-|\langle\alpha_\downarrow | \psi \rangle|^2$,
where $\alpha=1,2$ and $\left|\psi\right>$ is an edge state vector.
Taking $k=\pm 0.2$ and the lower branch of edge states in Fig.~3(b) as
an example, we show the spin-polarization value of the edge states in Fig.~3(c).
From Fig.~3(c) we see that the edge states of same $\mathbf{k}$ but opposite spins are located separately, 
resulting in finite spin-polarization.
A demonstration of dissipationless transport owing to the topological helical edge states is given in Sec.~C of
Supplementary Information.


Besides edge states, corner states also
emerge in a quadrupole phase, which is an essential signature that
differs the high-order TIs from conventional TIs.
For demonstrating the existences of corner states, we consider a finite
sample spanning $10\times10$ unit cells with open boundaries as
displayed in Fig.~4(a).
This sample has two types of edges and also two types of corners, i.e., zigzag and armchair. By solving its Hamiltonian, we observe eight corner states totally
as there are four corners and also two spins. Four of the corner states
have positive energies, and other four have negative energies.
Corner states localized at zigzag and armchair corners also have different energies.
For each corner state, there is $e/2$ amount of charge.
In Sec.~D of Supplementary Information the detailed charge distribution of corner states is given. As corner states have zero momentum, the spin-up
and spin-down corner states are degenerate due to time-reversal
symmetry. Thus, the corner states are not polarized along z-direction
of spins.
But due to finite spin-spin coupling, they are polarized along x- or
y-direction of spins. To see this, we redefine the spins by
$|\alpha_\uparrow^\prime\rangle,|\alpha_\downarrow^\prime\rangle=|\alpha_\uparrow\rangle\pm\text{i}|\alpha_\downarrow\rangle$ with $\alpha=1,2$.
Then taking the corner states with positive energies as an example, we calculate their spin-polarization according to this definition.
The result is displayed in Fig.~4(b). We see that for zigzag corners, the corner states are
spin-polarized up, whereas, for armchair ones, the states are
polarized down. The spin-polarization of each corner state is constant regardless of the values of
$\gamma$ and $\gamma^\prime$ as long as $|\gamma^\prime/\gamma|>1$, testifying their
topological nature. These spin-polarized topological corner
states may work as spinful quantum dots, with potential applications in spintronics.

In above discussions, we show the existence of topological helical edge states and also spin-polarized corner states in the
honeycomb model. Here we try to find a general condition for
emergence of helical edge states in a spinful quadrupole
phase. We denote the spinful quadrupole-induced localized states on the
edge $\zeta$ with momentum $\mathbf{k}$ and spin $\sigma=(\uparrow,\downarrow)$ as
$|\mathbf{k},\sigma,\zeta\rangle$. 
Here $\zeta(=L,R)$ indicates left-side edge ``L'' or right-side edge
``R'' of a ribbon.
To obtain spin-polarized
edge states, it is required that
$|\mathbf{k},\uparrow,\zeta\rangle$ and $|\mathbf{k},\downarrow,\zeta\rangle$
are not degenerate as shown in the honeycomb model. To fulfill this
condition, we check if there is any symmetry connecting these two
states. Here we consider three elementary symmetries such as time-reversal
$\mathcal{T}$, mirror reflections $\mathcal{M}_x$ and $\mathcal{M}_y$,
and $\pi$ rotation along z-direction $R_\pi$. Simply we have
\begin{align*}
   \mathcal{T}|\mathbf{k},\sigma,\zeta\rangle &=|-\mathbf{k},\bar{\sigma},\zeta\rangle,\\
   \mathcal{M}_x|\mathbf{k},\sigma,\zeta\rangle&=|\mathbf{k},\bar{\sigma},\bar{\zeta}\rangle,\\
   \mathcal{M}_y|\mathbf{k},\sigma,\zeta\rangle&=|-\mathbf{k},\bar{\sigma},\zeta\rangle,\\
   \mathcal{R}_\pi|\mathbf{k},\sigma,\zeta\rangle&=|-\mathbf{k},\sigma,\bar{\zeta}\rangle,
\end{align*}
where $\sigma$, $\bar{\sigma}$ have opposite values, and so as
$\zeta$, $\bar{\zeta}$.
From above relations, it is noticed that the above symmetric operations
change either two of these three ``quantum numbers''. Suppose that the
$|\mathbf{k},\uparrow,L\rangle$ state is $|0,0,0\rangle$ and the
$|\mathbf{k},\downarrow,L\rangle$ is $|0,1,0\rangle$, it seen that 
any single and combinations of these symmetric operations cannot
connect the two edge states as these symmetric operations
conserve the summation parity of these three ``quantum numbers''.
In other words, in a spinful quadrupole phase, the edge states are
spin-polarized in general, which is a quite unconventional result.

Finally, we discuss the relation of the proposed honeycomb model with conventional TIs that are supported by spin-orbital couplings.
In the honeycomb model, edge and corner states are protected by
finite charge polarization, which correspond to a winding
phase of a connection defined by Bloch functions in momentum space as
shown in Eq.~(1).
Such the nonzero winding phase can also be expressed as an integration
of a curvature by adding one extra dimension~\cite{Resta1994}. As discussed in Ref.~\cite{Qi2008}, by dimensional reduction a 2D
TI can be mapped to 1D SSH model. Thus, the proposed honeycomb model that is similar to 2D SSH model corresponds to a new type of 3D TI.
To see this, we investigate an adiabatic pumping process of the
spinful quadrupole in the honeycomb model controlled by parameter $\theta$. Namely, we set
$(\mu,\gamma,\gamma^\prime)=(\cos \theta, 0.2, \sin \theta),$ 
where $\mu$ is a staggered onsite potential with opposite signs on the
even- and odd-numbered atomic orbitals in a unit cell. The pumping
spectrum for half period of the finite sample of Fig.~4(a) is
displayed in Fig.~5(a). It is made up of three portion -- bulk, edge,
and corner states, which are colored as black, blue, and red in Fig.~5(a), respectively.
By replacing the pumping parameter $\theta$ with a quasi-momentum $k_z$ along the third
direction, we end up with a class of 3D TIs featuring spin-polarized surface and hinge states as shown in Fig.~5(b).
To realize this type of 3D TIs, one may stack the
honeycomb structure with
onsite potentials depending on the layer-index and a small inter-layer hopping.

We have discussed a spinful quadrupole phase as exemplified on a honeycomb lattice with Kekul\'{e}-like
hopping texture. With neither spin-orbital couplings nor external
fields, topological helical edge states closely resembling
those in conventional TIs have been created plus
spin-polarized corner states. By an adiabatic pumping process of spinful quadrupole, we have
defined a new class of three-dimensional topological insulators
characterized by spin-polarized surface and hinge states.
These results are expected to be useful for understanding the
topological properties of crystalline systems
and designing novel topological materials for low-power electronics.
\begin{figure}[h]
\begin{center}
\leavevmode
\includegraphics[clip=true,width=0.9\columnwidth]{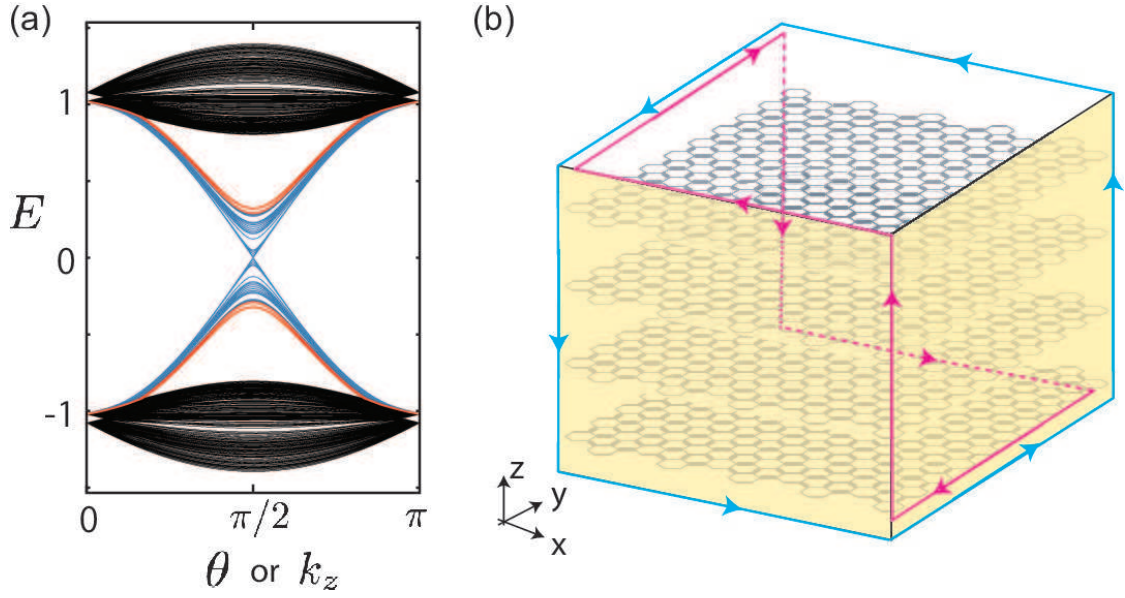}
\caption{(a) Pumping spectrum of spinful quadrupole for a half period. The spectrum consists
  of three types of states: the bulk states (black), edge states (blue)
  and corner states (red). (b) By replacing $\theta$ with a quasi-
  wavenumber $k_z$ along the third direction, a class of 3D
  topological insulators characterized by spin polarized hinge states --  blue for spin-up and red for spin-down -- emerge. }
\end{center}
\end{figure}

\clearpage

\newcounter{Cequ}
\newenvironment{Cequation}
   {\stepcounter{Cequ}%
     \addtocounter{equation}{-1}%
     \renewcommand\theequation{S\arabic{Cequ}}\equation}
   {\endequation}

\newcommand{\sbigotimes}{
  \mathop{\mathchoice{\textstyle\bigotimes}{\bigotimes}{\bigotimes}{\bigotimes}}
}

\setcounter{figure}{0}
\renewcommand{\thefigure}{S\arabic{figure}}
\section*{Supplementary A: Electric multipoles in crystalline systems}
In this section we consider a two-dimensional (2D) insulator to relate
the electric multipoles with Berry connection. The extension to 3D
systems is straightforward. 

The primitive vectors in the 2D insulator are denoted by $\mathbf{a}_1$ and
$\mathbf{a}_2$. A unit cell is labeled as 
$\mathbf{R}=n_1\mathbf{a}_1+n_2\mathbf{a}_2$, where $n_1$ and $n_2$ are
integers. The wave vector is given as 
$\mathbf{k}=\frac{1}{N}(n_1\mathbf{b}_1+n_2\mathbf{b}_2)$, where $\mathbf{b}_1$ and 
$\mathbf{b}_2$ are the primitive reciprocal lattice vectors defined by
$\mathbf{a_i\cdot b_j}=2\pi\delta_{ij}$ with $i,j=1,2$.
We assume that there are totally $N$ unit cells in the system and
each unit cell contains $N_{\rm orb}$ atomic orbitals. 
An atomic orbital $|\alpha\rangle$ located at $\mathbf{R}_\alpha\coloneqq \mathbf{R}+\mathbf{r}_\alpha$ is denoted by 
$|\mathbf{R},\alpha\rangle$. 
So then, we can express the Bloch states for $n$-th energy band as 
\begin{Cequation}
\left|{n\mathbf{k}}\right>=\frac{1}{\sqrt{N}}\sum_\mathbf{R} \sum_\alpha e^{i\mathbf{k}\cdot\mathbf{R}} u^\alpha_{n\mathbf{k}}\left|\mathbf{R}\alpha\right>.
\end{Cequation}
Here $u^\alpha_{n\mathbf{k}}$ is periodic part of Bloch function, which
is given by
\begin{Cequation}
\langle \mathbf{R}\alpha | n\mathbf{k}\rangle = \frac{1}{\sqrt{N}}
e^{i\mathbf{k}\cdot\mathbf{R}} u^\alpha_{n\mathbf{k}}.
\end{Cequation}


The dipole (per unit cell) is given by 
\begin{Cequation}
\mathbf{P} = \frac{e}{N} \sum^{N_{occ}}_{n}\sum_\mathbf{k} \langle n\mathbf{k}|\hat{\mathbf{x}}|n\mathbf{k}\rangle,
\end{Cequation}
where the summation over $n$ restricts to occupied energy bands labeled
with integer $n = 1, ..., N_{occ}$ and $\hat{\mathbf{x}}$ is the
position operator.
The symmetric quadrupole tensor $Q$ (per unit cell) has components $Q_{ij}$, which is given by
\begin{Cequation}
Q_{ij} = \frac{e}{N}\sum^{N_{occ}}_{n}\sum_\mathbf{k}\langle n\mathbf{k}|\hat{x}_i\hat{x}_j|n\mathbf{k}\rangle,
\end{Cequation} 
where $\hat{x}_i$ is the position operator along $i$-th direction

To find the eigenvalue of $\hat{\mathbf{x}}$, we look at the generator
of $\hat{\mathbf{x}}$ defined in the reciprocal lattice as
\begin{Cequation}
\hat{g}(\delta\mathbf{k}) = e^{i\delta\mathbf{k}\cdot\hat{\mathbf{x}}} = 
\sum_{\mathbf{R}}
\sum_{\alpha}
 e^{i\delta\mathbf{k}\cdot\mathbf{R}_\alpha}|\mathbf{R}\alpha\rangle\langle \mathbf{R}\alpha| .
\end{Cequation}
Here $\delta\mathbf{k} = \mathbf{n}~\delta k$, with $\mathbf{n}$ being a
unit vector. It is noted that this generator can be manipulated to
extract any order of the multipole, but here we only consider the dipole
and the quadrupole. 
This generator shifts the wavefunction of each energy band by $\delta
\mathbf{k}$ in the BZ. To obtain the geometric phase, we need to
adiabatically translate the wavefunction along a loop in the reciprocal space.
Thus, a geometric phase around a ``loop'' is related with 
eigenvalues of $\hat{\mathbf{x}}$ operator. 

\begin{figure}[!h]
\begin{center}
\leavevmode
\includegraphics[clip=true,width=0.4\columnwidth]{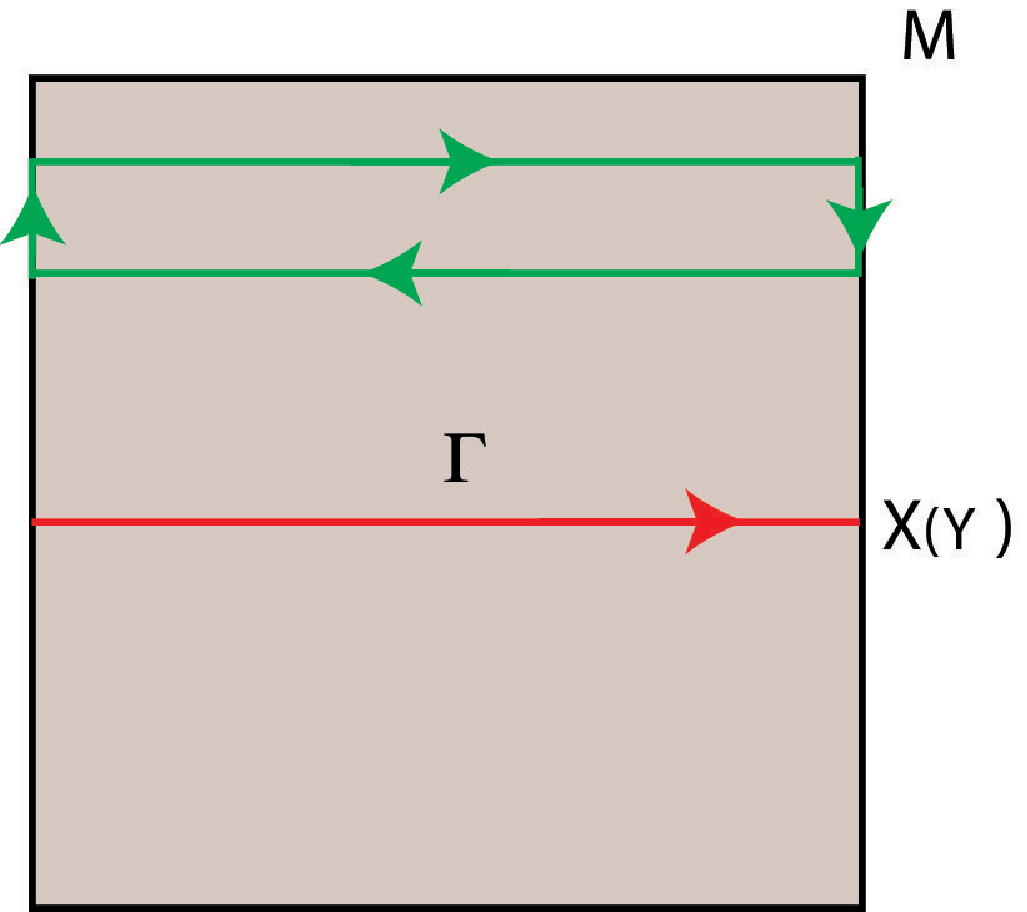}
\caption{Two types of loops in 2D BZ, where the red represents the first
 type of loop without direction changes, and the green represents the
 second type of loop where direction changes involve. Here we take
 square lattice as an example.} 
\end{center}
\end{figure}

To see this more clearly, let us consider two type of loops in the
2D Brillouin zone (BZ) as shown in Figure S1. 
First type of ``loop'' is a straight path that connects two equivalent
$\mathbf{k}$ points in the BZ, i.e. the red arrow in Figure S1. 
Since the path is connecting two equivalent
$\mathbf{k}$ points in the BZ, this path can be essentially viewed as a loop. 
This first type of ``loop'' corresponds to charge polarization and also 
electric quadrupole.
Second type of loop involves changes of directions in momemtum space,
i.e. the green arrow in Figure S1. 
 The second type of loops measures the change of charge
polarization to momemtum.

For a while, let us focus on the first type of ``loop'' to obtain the
charge polarization and electric quadrupole.
In this case, the unit vector $\mathbf{n}$ of the momentum shift is
parallel to the path. 
If we consider the loop which starts from 
$\mathbf{k}^L_1$ and ends at
$\mathbf{k}^L_{N_k}$, the set of momemntum on the loop $L$ is given as  
\begin{displaymath}
L\colon 
\mathbf{k}^L_1,
\mathbf{k}^L_2,
\mathbf{k}^L_3,
\cdots,
\mathbf{k}^L_{N_k-1},
\mathbf{k}^L_{N_k} (\coloneqq \mathbf{k}^L_1),
\end{displaymath}
where 
$\mathbf{k}^L_{i+1}-\mathbf{k}^L_{i}=\delta k \mathbf{n}$ and 
$N_k$ is total number of $\mathbf{k}$ points in the loop $L$.
Note $\mathbf{k}^L_1$ and $\mathbf{k}^L_{N_k}$ are equivalent points.


Next we project $\hat{g}$ unto the subspace of occupied energy bands on those $\mathbf{k}^L_i$ points forming the ``loop'' $L$ by the projection operator
\begin{displaymath}
\hat{P} = \sum^{N_{\rm occ}}_{n}\sum_{i}^{N_k} |n\mathbf{k}^L_i\rangle\langle n\mathbf{k}^L_i|
\end{displaymath} 
and call the projected generator $\hat{G}^L(\delta\mathbf{k})$, given by
\begin{Cequation}
\hat{G}^L(\delta\mathbf{k}) = \hat{P}\hat{g}(\delta\mathbf{k})\hat{P} =
 e^{i\delta\mathbf{k}\cdot \hat{\mathbf{X}}} =
 \sum_{m,n}^{N_{occ}}\sum_{i}^{N_k}\Gamma^{m,\mathbf{k}^L_i+\delta\mathbf{k}}_
					  {n,\mathbf{k}^L_i}
 |m,\mathbf{k}^L_i+\delta\mathbf{k}\rangle\langle
 n,\mathbf{k}^L_i|, \label{6} 
\end{Cequation}
where $\hat{\mathbf{X}} = \hat{P}\hat{\mathbf{x}}\hat{P}$ is the
projected position operator, and $
\Gamma^{m,\mathbf{k}_1}_{n,\mathbf{k}_2} = \langle
u_{m,\mathbf{k}_1}|u_{n,\mathbf{k}_2}\rangle$.
Relevant
to our current interest, we consider $\delta \mathbf{k}$ to be infinitesimal, and
thus  
\begin{Cequation}
\Gamma^{m,\mathbf{k}^L_i+\delta\mathbf{k}}_{n,\mathbf{k}^L_i} \approx
 \delta_{n,m} \Gamma^{m,\mathbf{k}^L_i+\delta\mathbf{k}}_{n,\mathbf{k}^L_i}.  
\end{Cequation}

As such, $\hat{G}^L(\delta\mathbf{k})$ can be reduced into a block diagonal form for each occupied energy band with each block given by
\begin{Cequation}
\hat{G}^{n,L}(\delta\mathbf{k}) =
 \sum_{i}^{N_k}
  \Gamma^{n,\mathbf{k}^L_i+\delta\mathbf{k}}_{n,\mathbf{k}^L_i}
 |n,\mathbf{k}^L_i+\delta\mathbf{k}\rangle\langle n,\mathbf{k}^L_i|, 
\end{Cequation} 
where $n$ is band index. Note that 
\begin{Cequation}
\Gamma^{n,\mathbf{k}^L+\delta\mathbf{k}}_{n,\mathbf{k}^L} = e^{i\delta\mathbf{k}\cdot \mathbf{A}^{n,n}(\mathbf{k}^L)},
\end{Cequation}
where $\mathbf{A}^{n,n}(\mathbf{k}) = i\langle u_{n,\mathbf{k}}|\partial_\mathbf{k}|u_{n,\mathbf{k}}\rangle$ is the Berry connection for the $n$th energy band. 

We now look for the eigenvalues of $\hat{G}(\delta\mathbf{k})$, which
are related to the eigenvalues of $\hat{\mathbf{X}}$, see
Eq. (\ref{6}). There are in total $N_k$ these values for each loop and let
them be $\mathbf{X}^{n,L}_j$, where $j = 1,...,N_k$.
Now we can write 
\begin{Cequation}
\hat{G}^{n,L}(\delta\mathbf{k}) = \sum_j e^{i\delta\mathbf{k}\cdot \mathbf{X}^{n,L}_j}|\mathbf{X}^{n,L}_j\rangle \langle \mathbf{X}^{n,L}_j|. \label{11}
\end{Cequation}
Here $|\mathbf{X}^{n,L}_j\rangle$ are the eigenvectors. This shows that $\hat{G}^{n,L}(\delta\mathbf{k})$ is unitary. 

Considering that $\hat{G}^{n,L}(\delta\mathbf{k})$ simply shifts vectors
along a loop, we can easily show that  
\begin{Cequation}
\left[\hat{G}^{n,L}(\delta\mathbf{k})\right]^{N_k}|\mathbf{k}^L_1\rangle 
=
 \left(\prod^{N_k}_{i=1}\Gamma^{n,\mathbf{k}^L_i+\delta\mathbf{k}}_{n,\mathbf{k}^L_i}\right)|\mathbf{k}^L_1\rangle 
 =
 e^{i\delta\mathbf{k}\cdot\sum_i\mathbf{A}^{n,n}(\mathbf{k}^L_i)}|\mathbf{k}^L_1\rangle
 = e^{i \int_L d\mathbf{k} \cdot
 \mathbf{A}^{n,n}(\mathbf{k})}|\mathbf{k}^L_1\rangle, \label{12} 
\end{Cequation}
where $\mathbf{k}^L_i$ denotes the $i$-th vector on the loop $L$ with
$i=1,...,N_k$. This shows that the eigenvalues of
$\left[\hat{G}^{n,L}(\delta\mathbf{k})\right]^{N_k}$ are 
$e^{i \int_L d\mathbf{k} \cdot \mathbf{A}^{n,n}(\mathbf{k})}$
independent of $i$, i.e. the starting point of loop.
On the other hand, from Eq.~(\ref{11}), we have 
\begin{Cequation}
\left[\hat{G}^{n,L}(\delta\mathbf{k})\right]^{N_k} = \sum_j
 e^{iN_k\delta\mathbf{k}\cdot
 \mathbf{X}^{n,L}_j}|\mathbf{X}^{n,L}_j\rangle \langle
 \mathbf{X}^{n,L}_j|. 
\end{Cequation}
Comparing this equation with Eq.~(\ref{12}), we immediately arrive at
\begin{Cequation}
N_k\delta\mathbf{k}\cdot \mathbf{X}^{n,L}_j =  \delta\mathbf{k} \cdot
 \sum_i\mathbf{A}^{n,n}(\mathbf{k}^L_i) + 2\pi M_j, \quad M_j = M_0,
 M_0+1, ..., M_0+N_k-1, \label{key} 
\end{Cequation}
where $M_0$ can be any integer and it does not have any physical
consequences, i.e. gauge freedom due to free choice of coordinate
origin. 
To make it more convenient to use, let us write 
\begin{displaymath}
\delta
\mathbf{k} = \frac{1}{N_k}(l_1\mathbf{b}_1+l_2\mathbf{b}_2), \quad
\mathbf{X}^{n,L}_{j} = \mathbf{R}^L_j + \mathbf{d}^{n,L}, \quad
\mathbf{R}^{L}_j = n_1\mathbf{a}_1+n_2\mathbf{a}_2,
\end{displaymath}
where $l_1$ and
$l_2$ are integers and so are $n_1$ and $n_2$. Substituting this in
Eq.~(\ref{key}), we obtain 
\begin{Cequation}
(l_1\mathbf{b}_1+l_2\mathbf{b}_2) \cdot \mathbf{d}^{n,L}=
 (l_1\mathbf{b}_1+l_2\mathbf{b}_2)\cdot
 \frac{1}{N_k}\sum_i\mathbf{A}^{n,n}(\mathbf{k}^L_i), \quad l_1n_1 +
 l_2n_2 = M_j.  
\end{Cequation}
We conclude that 
\begin{displaymath}
\mathbf{d}^{n,L} \cdot \mathbf{n}=
 {N_k}^{-1}\sum_i\mathbf{A}^{n,n}(\mathbf{k}^L_i)  \cdot \mathbf{n}.
\end{displaymath}
Since this relation must hold for any choice of $(l_1,l_2)$, we can set $l_1=1$ and $l_2=0$ without loss of generality. It follows that
\begin{Cequation}
\mathbf{d}^n(k_2)\cdot \mathbf{n} = \frac{1}{2\pi}\int^1_0 dk_1 \mathbf{A}^{n,n}(k_1\mathbf{b}_1 + k_2\mathbf{b}_2) \cdot \mathbf{n},\quad \mathbf{R}(j,j_0) = j\mathbf{a}_1+j_0\mathbf{a}_2.
\end{Cequation}
Here $k_2$ labels the loops and $(j,j_0)$ chooses the origin.

This result can now be used to calculate any multipole. Up to a constant term, we arrive at
\begin{Cequation}
\mathbf{P}(\mathbf{k})=e\sum_n^{N_{\rm occ}} \mathbf{d}^n(\mathbf{k}), \quad Q(\mathbf{k})= e\sum_n^{N_{\rm occ}} \mathbf{d}^n (\mathbf{k})\mathbf{d}^n(\mathbf{k}).
\end{Cequation}

\textbf{The case of zero Berry curvature.} 
Here we assume the Berry curvature given by $\mathcal{F}=\nabla_\mathbf{k} \times \mathbf{A}$ equals to zero at any $\mathbf{k}$ points. In this case, any integrations of $\mathbf{A}$ along a second type of loop are zero, and $\mathbf{P}$ and $Q$ become independent of $\mathbf{k}$. To see this, we can consider a second type of loop as shown by the green loop in Figure~S1. Starting from the left-bottom and along the clock-wise direction, the four joint points of the loop are $(k_i,k_1)$, $(k_i,k_2)$, $(k_{N_L},k_2)$ and $(k_{N_L},k_1)$, accordingly. Then the integration of $\mathbf{A}$ over the loop is given by
\begin{displaymath}
\begin{split}
\oint d\mathbf{k}\cdot\mathbf{A}&=\int_{k_1}^{k_2} dk^\prime A_2(k_i,k^\prime)+\int_{k_2}^{k_1} dk^\prime A_2(k_{N_L},k^\prime)\\
&\quad +\int_{-\mathbf{b}_1/2}^{\mathbf{b}_1/2} dk A_1(k,k_2)+\int_{\mathbf{b}_1/2}^{-\mathbf{b}_1/2} dk A_1(k,k_1)\\
&=d_1(k_2)-d_1(k_1) \\
&=\iint \mathcal{F} ds\\
&=0,
\end{split}
\end{displaymath}

where the first two terms cancel as $k_i$ and $k_{N_L}$ are two equivalent points, $A_i=\mathbf{A}\cdot\mathbf{n}_i$ and $d_1=\mathbf{d}\cdot\mathbf{n}_1$. Then we have $d_1(k_1)=d_1(k_2)$ regardless of $k_1$ and $k_2$. Similar argument can also be applied to $d_2$. As a result of zero Berry curvature, $\mathbf{P}$ and $Q$ become constants in BZ, and their expressions can be further simplified. We arrive at the expressions of dipole and quadrupole in the main text as
\begin{Cequation}
\mathbf{P}_i=e\sum_n^{N_{occ}}(\mathbf{d}^n\cdot\mathbf{n}_i),\quad Q_{ij}=e\sum_n^{N_{occ}}(\mathbf{n}_i\cdot\mathbf{d}^n)(\mathbf{n}_j\cdot\mathbf{d}^n),
\end{Cequation}
where $\mathbf{d}^n\cdot \mathbf{n}=\frac{1}{2\pi}\int_L d\mathbf{k}\cdot\mathbf{A}^{n,n}$ with $L$ a first type of loop that connects two equivalent points.

\section*{Supplementary B: Energy bands structure of honeycomb 2D SSH model}
\begin{figure}[!h]
\begin{center}
\leavevmode
\includegraphics[clip=true,width=1.0\columnwidth]{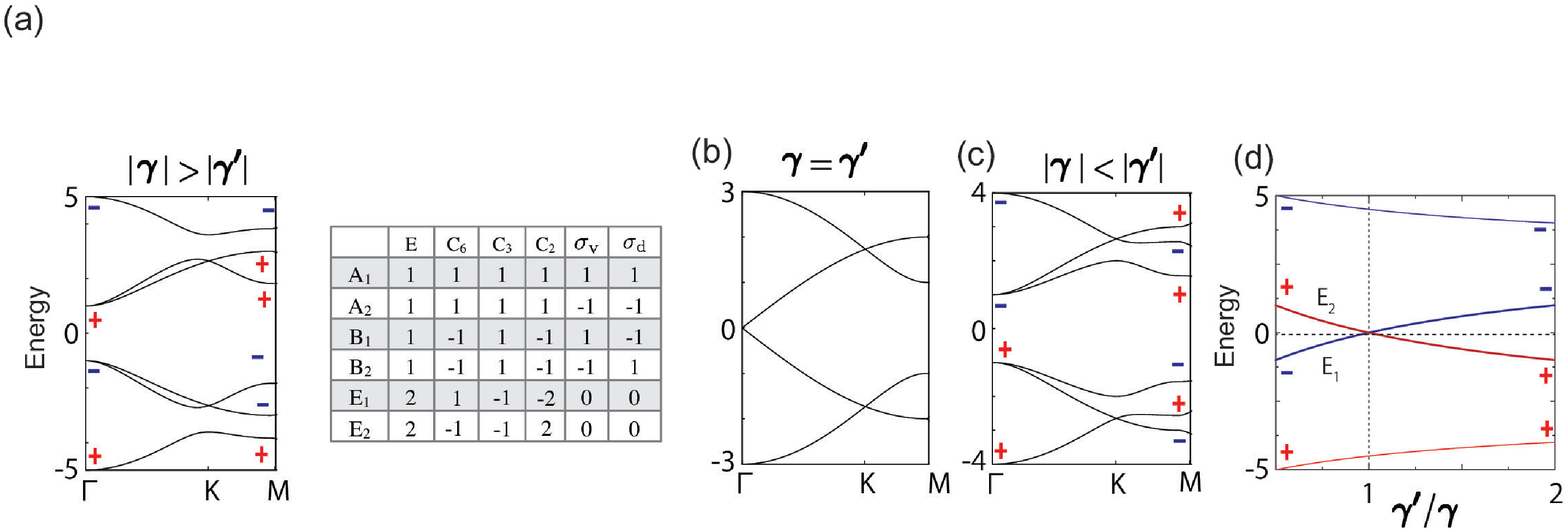}
\caption{(a) Energy band structure of 2D honeycomb SSH model for $\gamma=-2.0$ and $\gamma=-1.0$. Inset is the character table of $C_{6v}$ point group symmetry. (b) Energy band structure at $\gamma=\gamma^\prime$. (c) Energy band structure for $\gamma=-1.0$ and $\gamma^\prime=-2.0$. (d) energy band evolution at $\Gamma$ point for $\gamma^\prime/\gamma$. The parities at $\Gamma$ and M points are marked as $\pm$.}
\end{center}
\end{figure}
The full Hamiltonian of 2D SSH model sitting on a honeycomb lattice is given by\cite{Liu2017B}

\begin{Cequation}
\mathcal{H}=
\begin{pmatrix}
0&e^{\text{i}\phi_3}&0&te^{\text{i}\phi_1}&0&e^{\text{i}\phi_2}\\

e^{-\text{i}\phi_3}&0&e^{-\text{i}\phi_1}&0&te^{-\text{i}\phi_2}&0\\

0&e^{\text{i}\phi_1}&0&te^{\text{i}\phi_2}&0&e^{t\text{i}\phi_2}\\

te^{-\text{i}\phi_1}&0&e^{-\text{i}\phi_2}&0&e^{-\text{i}\phi_3}&0\\

0&te^{\text{i}\phi_2}&0&e^{\text{i}\phi_3}&0&e^{\text{i}\phi_1}\\

e^{-\text{i}\phi_2}&0&et^{-\text{i}\phi_3}&0&e^{-\text{i}\phi_1}&0

\end{pmatrix},
\end{Cequation}
where the bases are the atomic sites $(|1\rangle, |2\rangle, |3\rangle, |4\rangle, |5\rangle, |6\rangle)^T$, which are shown in the main text of Fig.~1\textbf{a}, and $\phi_i$ is the Bloch's phase given by $\mathbf{k}\cdot\mathbf{r}_i$ with
\begin{Cequation}
\mathbf{r}_1=a(0,1/\sqrt{3}),\text{  }\mathbf{r}_2=a(-1/2,-1/2\sqrt{3}),\text{  }\mathbf{r}_3=a(1/2,-1/2\sqrt{3}).    
\end{Cequation}

The energy spectrum for $\gamma=-2.0$ and $\gamma^\prime=-1.0$ is plotted in Fig.~S2(a). There are six energy bands based on the character table of C$_\text{6v}$ point group symmetry (inset of Fig.~S2a). The doublets energy bands coincide at $\Gamma$ point when $|\gamma|=|\gamma^\prime|$ as displayed in Fig.~S2(b), and a band inversion at $\Gamma$ point happens for $|\gamma^\prime|>|\gamma|$. In Fig.~S2(c) is displayed the enerby spectrum for $\gamma=-1.0$ and $\gamma^\prime=-2.0$. In Fig.~S2(d) the energy evolution of two doublets band at $\Gamma$ point for $\gamma^\prime/\gamma$ is displayed. The $\pm$ sign in Fig.~S2 indicates the eigenvalue of $C_2$ rotation of the wavefunction.
 
For more details about the Hamiltonian near $\Gamma$, we apply the following unitary matrix
\begin{Cequation}
\mathcal{U}=
\begin{pmatrix}
1&1&1&1&1&1\\
1&-1&1&-1&1&-1\\
1 & \omega^2 & \omega & 1 &\omega^2 &\omega \\
1 & \rho^5 &\omega^2 & -1 & \omega & \rho \\
1 & \omega &\omega^2 & 1 &\omega & \omega^2 \\
1 & \rho & \rho^2 & -1 & \omega^2 & \rho^5 \\
\end{pmatrix},
\end{Cequation}
where $\rho=e^{\text{i}\pi/3}$, $\omega=e^{\text{i}2\pi/3}$. From the first row to the last row, they correspond the two singlets energy bands, $|\downarrow_2\rangle$, $|\downarrow_1\rangle$, $|\uparrow_2\rangle$ and $|\uparrow_1\rangle$, respectively.

After the unitary transformation and expanding around $\Gamma$ point, we obtain 

\begin{Cequation}
\mathcal{H}=
\begin{pmatrix}
m_1 &0&0&m_2(k_x-ik_y)&0&-m_2(k_x+ik_y)\\
0&-m_1&m_2(k_x+ik_y)&0&-m_2(k_x-ik_y)&0\\
0&m_2(k_x-ik_y)&m_2&-m_1(k_x+ik_y)&0&0\\
m_2(k_x+ik_y)&0&-m_1(k_x-ik_y)&-m_2&0&0\\
0&-m_2(k_x+ik_y)&0&0&m_2&m_1(k_x-ik_y)\\
-m_2(k_x-ik_y)&0&0&0&m_1(k_x+ik_y)&-m_2\\
\end{pmatrix},
\end{Cequation}

where $m_1=2+t$, $m_2=-1+t$, and the bases are $(|A_1\rangle, |B_1\rangle, |\downarrow_2\rangle, |\downarrow_1\rangle, |\uparrow_2\rangle, |\uparrow_1\rangle)^T$, where $|A_1\rangle, |B_1\rangle$ are two singlet energy bands. From above Hamiltonian it is clear that one cannot just extract the doublets bands by throwing singlets bands away as the coupling among them is not zero and not topological trivial, which is $m_2(k_x \pm ik_y)$. Furthermore we can see that unlike conventional topological insulators, the spin-orbital-coupling-like terms $k_x+ik_y$ and $k_x-ik_y$ appear simultaneously for each atomic orbital.

\section*{Supplement C: Transport simulation of helical edge states}
\begin{figure}[!htp]
\begin{center}
\leavevmode
\includegraphics[clip=true,width=1.0\columnwidth]{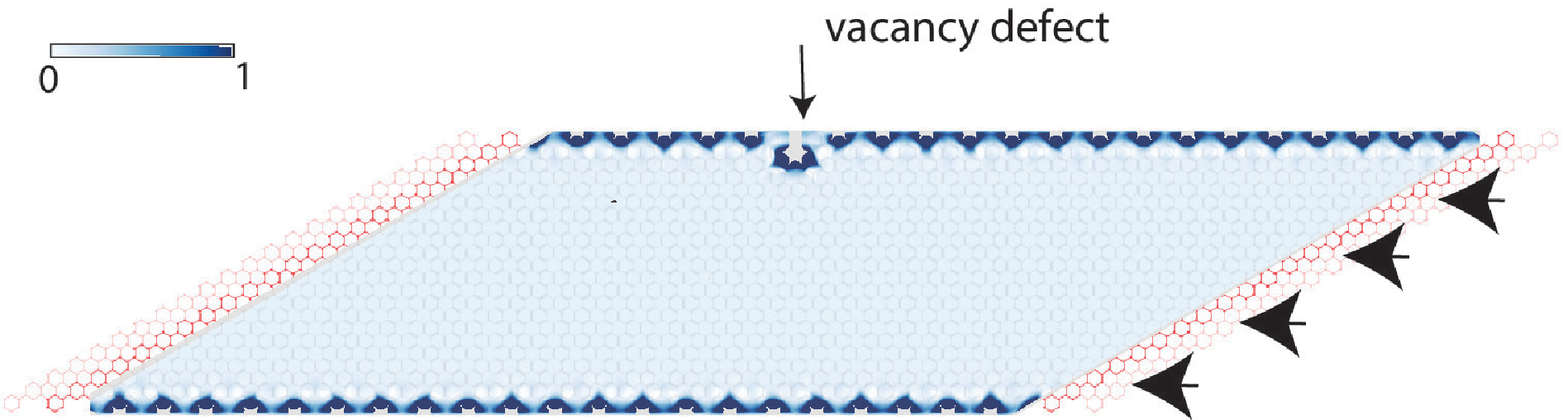}
\caption{Current is sent through attached leads made up by the same material (indicated by red) from right to left at $E\sim 0$. There is a vacancy defect in the upper edge. Helical topological edge state turns around the vacancy defect near the edge without back-scatterings. }
\end{center}
\end{figure}
Because of the mirror symmetry and finite spin-spin coupling, the topological edge states discussed in the main text are also helical. When the spin-spin coupling is small, these helical edge states are immune to back-scatterings. As displayed in Fig.~S3, the helical topological edge state induced by attached leads turns around a vacancy near the edge without apparent back-scatterings \cite{Kwant}.

\begin{figure}[!htp]
\begin{center}
\leavevmode
\includegraphics[clip=true,width=1.0\columnwidth]{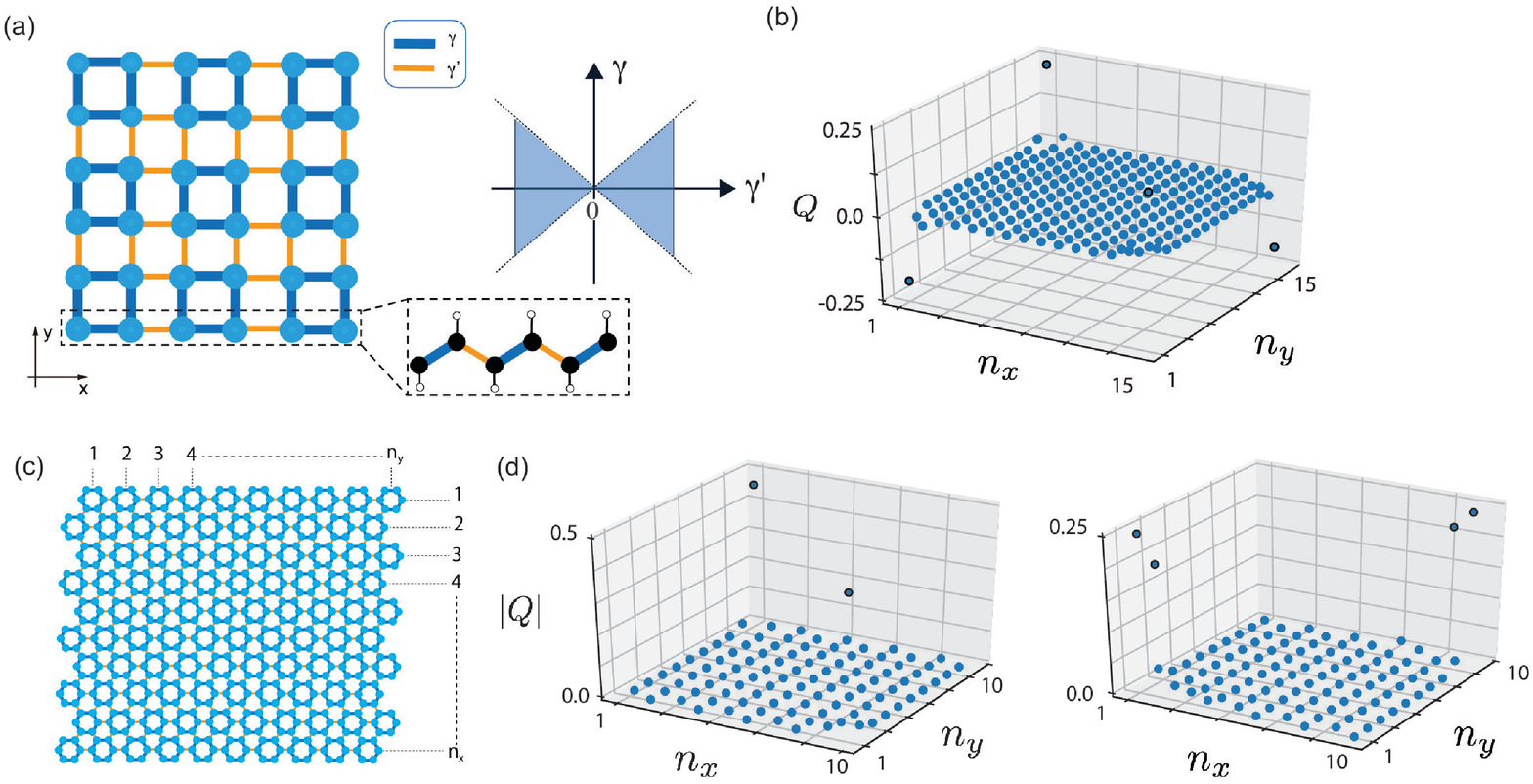}
\caption{(a) Schematic of 2D SSH model. The model is specified by an intra-cell hopping parameter $\gamma$ and an inter-cell one $\gamma'$. Two topologically distinct phases exist depending on whether $|\gamma|>|\gamma'|$ or not, as displayed in the diagram in the inset. In the shaded region, a topological dipole of $e/2$ and a topological quadrupole of $e/4$ coexist. (b) Fractional charge states localized at the corners demonstrated for a sample spanning $15\times 15$ unit cells. (c) Schematic of a sample comprised of $10\times10$ unit cells supporting the corner states. We call the edge along the $x$-direction zigzag, while that along the $y$-direction armchair. (d) Charge distribution in the corner states for $\gamma = -1.0$ and $\gamma' = -3.0$. For each state, the charges total to $e/2$.}
\end{center}
\end{figure}

\section*{Supplementary D: Fractional corner charge}

In this section we display the numerical results of localized fractional charge on the corners of 2D SSH (Su-Schrieffer-Heeger) model on both square and honeycomb lattices. 
SSH model is an intriguing system realizing the topological quadrupole phase. In Fig.~S4(a) the hopping texture of 2D SSH on a square lattice is displayed.  There are two types of hoppings, namely the intra-cell hopping $\gamma$ and the inter-cell hopping $\gamma^\prime$. Inset of Fig.~S4(a) displays its topological phase diagram in terms of intra-cell and inter-cell hoppings $\gamma$ and $\gamma^\prime$. Two topologically distinct phases can be identified depending on the ratio $t=\gamma^\prime/\gamma$.  For $t>1$, the topologically non-trivial phase results with $|P_x|=|P_y|=e/2$ and $|Q_{xy}|=e/4$ for the lowest band. Upon being terminated, topological edge and corner states appear, as demonstrated in Fig.~S4(b) for a sample spanning $15\times15$ unit cells. Unlike in the work of Benalcazar \textit{et al.}~\cite{Benalcazar2017, Benalcazar2017B}, there is no $\pi$-flux threading the unit cells in our system. As such, finite $P_i$ band $Q_{xy}$ coexist in a mixture state, which has no classical counterpart.

Similar to 2D SSH model on a square lattice,  topological corner states appear when $|\gamma^\prime|>|\gamma|$ in a honeycomb lattice. In Fig.~S4(c) the finite sample of honeycomb SSH model with open boundary conditions is displayed, which possesses both zigzag and armchair edges. In Fig.~S4(d) the fractional corner charge is displayed. Different from square lattice case, the fractional charge is $\pm e/2$ in honeycomb lattice as there are both pseudo-spin up and down contributions.

\bibliography{references}

\begin{thebibliography}{34}
\bibitem [{ons()}]{onsites}%
  \BibitemOpen
  \href@noop {} {}\bibinfo {note} {In this honeycomb model, such the zero Berry
  curvature condition can be fulfilled by adding tiny unsymmetric onsite
  potentials, which have negligible effects on the properties we are interested
  in. For example, one can put $(0,0.01,0.02,0,0.01,0.02)$ onsite potentials
  for 1st to 6th atomic orbitals as indexed in Fig.~1(a). Then because of
  simultaneous presences of time-reversal symmetry and inverison symmetry, the
  Berry curvature vanishes everywhere in momentum space.}\BibitemShut {Stop}%
\end{thebibliography}

\end{document}